# Independent and coherent transitions between antiferromagnetic states of few-molecule systems

Claire Besson,[a,b] Philipp Stegmann,[c,d] Michael Schnee,[b] Zeila Zanolli,[e,f,g] Simona Achilli,[f,g,h] Nils Wittemeier,[g] Asmus Vierck,[i] Robert Frielinghaus,[b] Paul Kögerler,[i,j] Janina Maultzsch,[i,k] Pablo Ordejón,[g] Claus M. Schneider,[b] Alfred Hucht,[c] Jürgen König,[c] Carola Meyer[j,b]*

a. Department of Chemistry, The George Washington University, Washington DC 20052, USA

b. Peter Grünberg Institut (PGI-6), Forschungszentrum Jülich, 52425 Jülich, Germany and Jülich Aachen Research Alliance (JARA)–Fundamentals of Future Information Technology, 52425 Jülich, Germany

c. Theoretische Physik, Universität Duisburg-Essen and CENIDE, 47048 Duisburg, Germany

d. Current affiliation: Department of Chemistry, Massachusetts Institute of Technology, Cambridge, Massachusetts 02139, USA

e. Condensed Matter and Interfaces, Debye Institute for Nanomaterials Science, Utrecht University, Princetonplein 1, 3584 CC Utrecht, The Netherlands

f. European Theoretical Spectroscopy Facility (ETSF)

g. Catalan Institute of Nanoscience and Nanotechnology - ICN2 (CSIC- BIST) Campus UAB, Bellaterra, 08193 Barcelona, Spain

h. Dipartimento di Fisica "Aldo Pontremoli", Universitá degli Studi di Milano, Via Celoria 16, Milan, Italy

i. Institut für Festkörperphysik, Technische Universität Berlin, Hardenbergstrasse 36, 10623 Berlin, Germany

j. Institute of Inorganic Chemistry, RWTH Aachen University, 52074 Aachen, Germany

k. Department of Physics, Friedrich-Alexander University Erlangen-Nürnberg, Staudtstrasse 7, 91058 Erlangen, Germany

l. School of Mathematics/Computer Sciences/Physics, Institute of Physics, Universität Osnabrück, D-49069 Osnabrück, Germany

* carola.meyer@uni-osnabrueck.de

ABSTRACT Spin-electronic devices are poised to become part of mainstream microelectronic technology. Downsizing them led to the field of molecular spintronics. Here, we provide proof of concept data that allows expanding this area from its traditional focus on single-molecule magnets to molecules in which spin centers are antiferromagnetically coupled to result in a singlet ground state. In this context, and in contrast to all previous work on molecular spintronics, we develop a detection scheme of the molecule's spin state that does not rely on a magnetic moment. Instead, we use quantum dot devices consisting of an isolated, contacted single-wall carbon nanotube covalently bound to a limited number of molecular antiferromagnets, for which we chose representative coordination complexes incorporating four Mn(II) or Co(II) ions. Time-dependent quantum transport measurements along the functionalized nanotube show step-like transitions between several distinct current levels that we attribute to transitions between different antiferromagnetic states of individual molecular complexes grafted on the nanotube. A statistical analysis of the switching events using factorial cumulants indicates that the cobalt complexes switch independently from each other whereas a coherent superposition of the antiferromagnetic spin states of the molecules along the nanotube is observed for the manganese complexes. The long coherence time of the superposition state (several seconds at 100 mK) is made possible by the absence of spin and orbital momentum in the relevant states of the manganese complex, while the cobalt complex includes a significant orbital momentum contribution due to the pseudo-octahedral coordination environment of the $d^7$ metal centers.

TEXT

Antiferromagnetic spintronics is emerging as a promising field for technologies ranging from magnetic random access memories to neuromorphic computing and THz information devices, as antiferromagnets (AFMs) do not produce any, and are insensitive towards, magnetic stray fields.[1-4] The lack of a net magnetic moment confers high stability to the antiferromagnetic states but, at the same time, makes it challenging to address and differentiate among them. Therefore, this area has not yet been extended to the level of individual molecules exhibiting antiferromagnetically coupled magnetic spin centers, while spintronics with e.g. high-spin single-molecule magnets (SMMs) are by now well-established.[5,6] In this article, we showcase how transitions between distinct molecular states with $S_{tot} = 0$ can be monitored via charge transport when the molecules are coupled to a carbon nanotube quantum dot in field effect geometry. Transitions between different antiferromagnetic states are accompanied by step-like changes of the current through the quantum dot, resulting in a random telegraph signal (RTS). We analyze the signal statistics employing factorial cumulants, which are well-established in the field of quantum optics but have not yet been widely applied in transport. This method of investigation paves the way to explore the molecular equivalent of AFM spintronics by employing states that do not couple to fluctuations of external magnetic fields, thereby allowing for long-lived superposition states. In this work we study two structurally identical molecular systems that differ fundamentally only in the spin carrying metal ions (Figure 1). Modelling the RTS, we find that in the absence of spin-orbit coupling such states can exhibit a coherent superposition with exceptionally long coherence time.

In the past years significant progress has allowed detecting spin states of individual molecules using scanning tunneling microscopy[7-17] or molecules in junctions.[18-21] However, in these approaches the electric bias field applied to the molecule will in general influence its charge state, and therefore its magnetic characteristics. To circumvent this issue, it has been proposed to detect the spin state of an SMM by grafting it to a non-magnetic one-dimensional conductor (such as a carbon nanotube, CNT) and measuring the magnetic state of the so-obtained hybrid system.[22] CNTs have been used successfully to read out states of SMMs by measuring the magnetoresistance effect,[23] detecting phonon-spin interaction[24] and spin interactions within an individual SMM.[25] All these experiments depend on the detection of a local magnetic moment. Therefore, they cannot be employed for single-molecule antiferromagnets that have no resulting magnetic moment. Here, we show that devices functionalized with antiferromagnetic molecules that differ only in the spin carrying metal ions exhibit random telegraph signals with distinct statistics that reveal whether the transitions between different current levels are correlated. We interpret the RTS in terms of transitions between different molecular $S_{tot} = 0$ eigenstates.

## Results and discussion

### Electronic and magnetic structure of the carbon nanotube-antiferromagnet hybrid

To fabricate a system that exhibits distinct molecular $S_{tot} = 0$ states, we use charge-neutral molecular $[M_4L_2(OAc)_4]$ complexes[26,27] ($H_2L$ = 2,6-bis-(1-(2-hydroxyphenyl)iminoethyl)-pyridine; HOAc = acetic acid; M = $Mn^{II}$ or $Co^{II}$). Those complexes (hereafter denoted as **{Mn₄}** and **{Co₄}**) are nearly isostructural and display a quasi-tetrahedral core of divalent metal ions interlinked by ligands (inset Figure 1). Exchange coupling between the metal ions is mediated mainly by four oxygen atoms of the ligands, each bridging between three metal centers into a nearly cubic $M_4O_4$ core, which can be thought of as two pairs of antiferromagnetically interacting ions. Two of the four ions reside in an approximately octahedral, two in an approximately trigonal bipyramidal ligand field with a spin $S = 5/2$ ($S = 3/2$) for $Mn^{II}$ ($Co^{II}$). The ground state is antiferromagnetic as supported by magnetization measurements for the pristine molecules as well as for molecules grafted to CNTs.[26-28] Density Functional Theory (DFT) calculations of the isolated complexes[27] also yield an $S_{tot}^z = 0$ ground state, confirming the overall antiferromagnetic nature of the intramolecular magnetic interactions in the simpler one-electron approximation. The complexes retain their antiferromagnetic nature when grafted covalently to the CNT (Figure **1**) by a metathesis reaction between an acetate ligand of the complex and a carboxylic acid group generated by oxidation of the tube.[28] Our DFT calculations confirm the overall antiferromagnetic nature of the ground state of the grafted complexes and demonstrate that the interaction between each individual molecule and the CNT is similar for **{Co₄}** and **{Mn₄}**: the electronic properties of the nanotube around the Fermi energy are slightly perturbed by the molecule, and the total spin polarization of the hybrid system is larger for **{Co₄}**-CNT than for **{Mn₄}**-CNT (SI-2). We used an open-system approach[29-31] to demonstrate that the spin polarization induced by magnetic exchange interaction between the spins of the molecule and the spin of the electron in the CNT quantum dot at the grafting site is long-ranged (Figure 1, SI-2).

Note that spin configurations with different total spin ($S_{tot} > 0$) lie several meV above the ground state for both complexes (Tables S1 and S2, SI-2). This is at least two orders of magnitude larger than the energy involved in the low-temperature transport experiments presented here.

Instead, we have to think of the AF ground states with $S_{\text{tot}} = S_{\text{tot}}^z = 0$ in terms of a total spin of the complex $\vec{S}_{\text{tot}} := \sum_i \vec{S}_i$ characterized by the spin quantum number $S_{\text{tot}}$ and the magnetic quantum number $S_{\text{tot}}^z$. To form a complete set of quantum numbers, we (arbitrarily) group the four individual ion spins into two pairs and include the spin quantum numbers $S_{12}$, $S_{34}$ of the two spin pairs $\vec{S}_{ij} := \vec{S}_i + \vec{S}_j$ so that $\vec{S}_{\text{tot}} := \vec{S}_{12} + \vec{S}_{34}$ (Figure S1 and supporting information section SI-1). We assume a Heisenberg spin-exchange coupling for each pair of spins, $H = -\frac{1}{2}\sum_{i \neq j} J_{ij}\vec{S}_i \cdot \vec{S}_j$ with antiferromagnetic $J_{ij} < 0$. If all exchange couplings are equal, $J_{ij} = J$, the eigenenergy $E = -\frac{J}{2}[S_{\text{tot}}(S_{\text{tot}} + 1) - 4S(S + 1)]$ only depends on the spin quantum number $S_{\text{tot}}$ of the total spin but not on $S_{\text{tot}}^z$, $S_{12}$, or $S_{34}$. For antiferromagnetic coupling, $J < 0$, the ground state has total spin $S_{\text{tot}} = 0$ with vanishing dipole moment, $S_{\text{tot}}^z = 0$, exhibiting a six-fold (four-fold) degeneracy for **{Mn₄}** (**{Co₄}**). The four degenerate states for **{Co₄}** would correspond to $|S_{\text{tot}}, S_{\text{tot}}^z, S_{12}, S_{34}\rangle = |0,0,0,0\rangle, |0,0,1,1\rangle, |0,0,2,2\rangle, |0,0,3,3\rangle$, to which the $|0,0,4,4\rangle$ and $|0,0,5,5\rangle$ states would be added for **{Mn₄}** (SI-1). However, the degeneracy is lost in the actual molecules because the exchange interactions $J_{ij}$ are not equivalent. This enables low energy transitions within the now split $S_{\text{tot}} = S_{\text{tot}}^z = 0$ manifold (an example for an excited state is given in SI-1) that cause a random telegraph signal in the current through the CNT quantum dot.

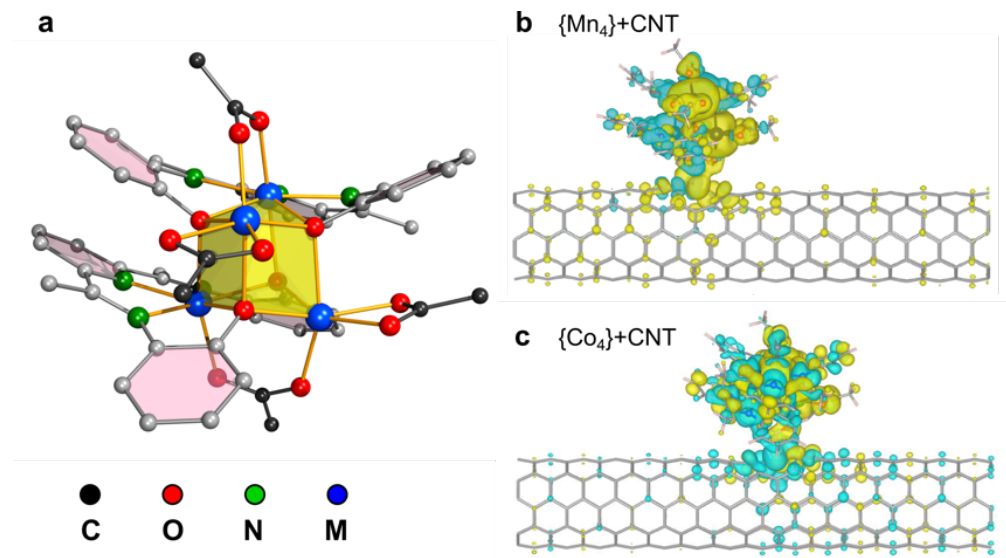

**Figure 1 a** General structure of the metal complexes (M: metal ion). H is omitted for clarity, the central (distorted) $M_4O_4$ cubane highlighted in yellow, all coordinative bonds in orange. All aromatic rings are made transparent to improve spatial representation. Ligand backbones differentiate between chelate and acetate ligands. **b,c** Spin density ($\rho_\uparrow - \rho_\downarrow$) on the {Mn$_4$}-CNT (b), and the {Co$_4$}-CNT (c) hybrid system computed from first principles in an open-system setup.[29-31] Light blue and yellow indicate positive and negative values of the isosurfaces normalized to the maximum spin density.

A key difference between the two molecules is the value of the orbital angular momentum ⟨L⟩, which is finite in Co$^{II}$, in agreement with the 3d$^7$ atomic configuration and the respective ligand fields, but almost negligible in Mn$^{II}$ due to the half filling of the 3d shell. By including spin orbit coupling in the DFT calculations we verified that the orbital angular momentum of cobalt is practically fully quenched for the two trigonal bipyramidal metal centers, while a significant orbital moment (0.18 and 0.25 $\mu_B$, open-system approach) is found for the two octahedrally coordinated atoms (SI-2).

**Device preparation and characterization**

Individual CNTs functionalized with either **{Mn₄}** or **{Co₄}** complexes are contacted in a field effect transistor structure with highly doped silicon serving as back gate. Basic device characterization and fabrication details for the {Mn₄}-based device have been published elsewhere.[32] A similar procedure was used for {Co₄}-based device fabrication. The number of complexes within a device is derived from our previous experimental evaluation[28] of the functionalization density – approximately one complex every 10 nm, meaning negligible through-space interaction between neighboring complexes. Analysis of Raman spectroscopy based on the G-mode[32,33] and on the TO+ZA combination mode[34] of the device with **{Mn₄}** complexes suggests a zigzag (15,0) metallic CNT (Figure S4, SI-3). Accordingly, no transport gap is found in the electrical measurements. This is consistent with our first-principles prediction of a metallic character for the hybrid CNT-**{M₄}** systems (Figure S2 and section SI-2). The stability diagrams of the quantum dots bearing about 50 **{Mn₄}** complexes or 70 **{Co₄}** complexes exhibit regular Coulomb diamonds (Figure 2). Figure 2a shows the stability diagram of the **{Mn₄}**-functionalized quantum dot measured at $T = 4$ K. Since excited states are not resolved, we use a charging energy of $E_{ch} \sim 4$ eV to get a rough estimate of the length of the CNT quantum dot with $l \sim 350$ nm.[32,35] The stability diagram of the quantum dot functionalized with **{Co₄}** complexes was measured at base temperature of the dilution refrigerator ($T \approx 100$ mK, Fig. 2b). Here, excited states are nicely resolved and thus a more accurate estimate of the quantum dot length is derived from the first excited state at $\Delta E = 0.7$ meV yielding $l = 714$ nm. The length estimates from the stability diagram fit reasonably well with the designed distance between the electrodes of 500 nm for the {Mn₄} device, and of 700 nm for {Co₄} device. We thus conclude that the covalent grafting of complexes to the CNT does not cause strong localization within the devices.

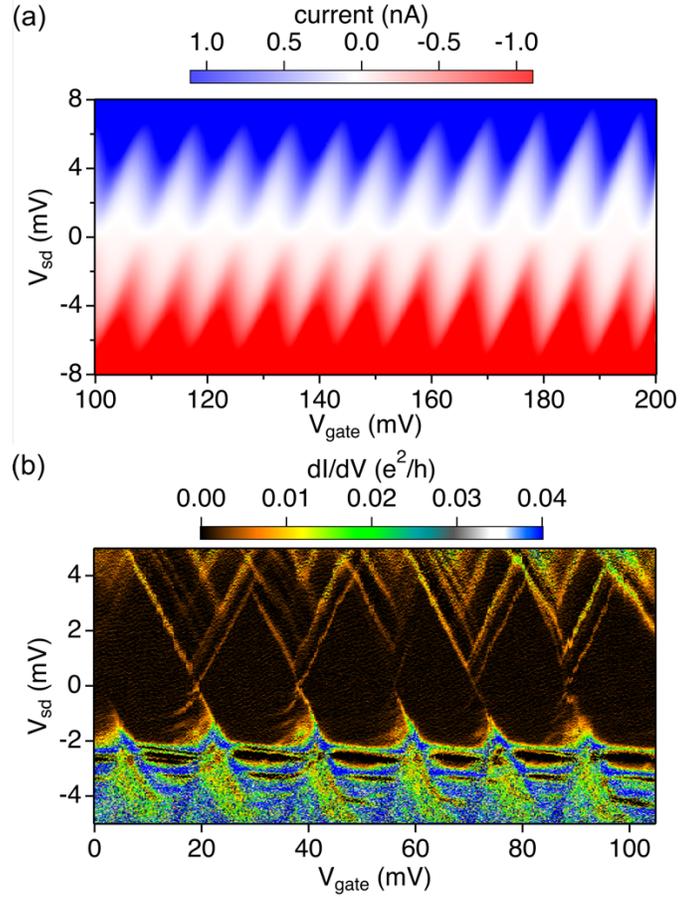

**Figure 2. a**: Stability diagram of the quantum dot formed on the **{Mn₄}**-functionalized CNT, measured at a temperature of 4 K ($B = 0$ T), exhibiting a very clean and regular pattern of Coulomb diamonds with a quantum dot size comparable to the distance between the electrodes.[32] **b**: Coulomb diamonds of a CNT quantum dot functionalized with **{Co₄}** complexes at 100 mK ($B = 0$ T). The width of the current steps as well as that of the Coulomb peaks yield an electron temperature of 600 mK ($\sim$ 50 µeV).

**Analysis of the random telegraph signals**

Figure 3a shows on the left a short section of the current trace taken at the base temperature of the dilution refrigerator at the edge of the Coulomb diamond of a **{Mn₄}** functionalized CNT device, displaying a characteristic random telegraph signal. The probability density function (PDF) corresponding to the entire trace (one hour length), reported in Figure 3a, exhibits a quasi-exponential decay instead of a single Gaussian peak expected for simple noise. A similar signal is measured at an equivalent position of a CNT quantum dot functionalized with {**Co₄**} complexes (Figure 4). While it does clearly not resemble a single Gaussian peak, the asymmetry of the PDF is much less prominent compared to the {**Mn₄**} device and the signal decays much faster, indicating the presence of a smaller number of levels for the signal.

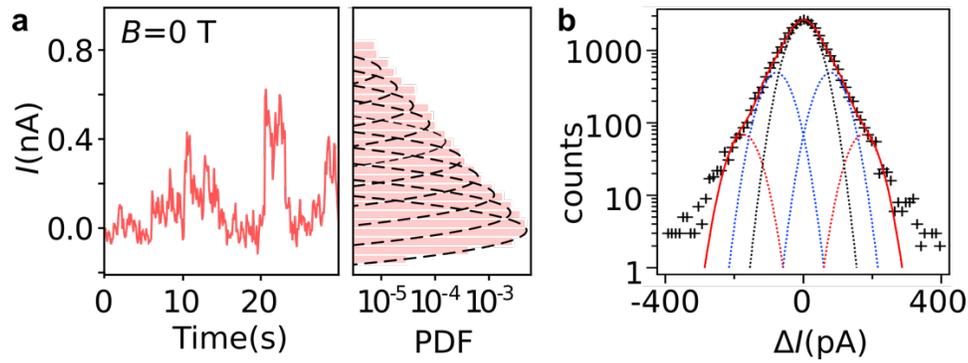

**Figure 3. a**: Short section of the time trace of the current at the edge of a Coulomb diamond of the {**Mn₄**}-functionalized CNT quantum dot and PDF of the entire one hour signal displayed with 11 Gaussian peaks indicating the current levels (we omit the 3 highest levels for better visibility). **b**: Histogram of the derivative of the entire signal (note the logarithmic scale) fitted with five Gaussian peaks leading to a standard deviation of the direct signal $\sigma_{direct} = \sigma_{deriv} / \sqrt{2} = 40(1)$ pA respectively.

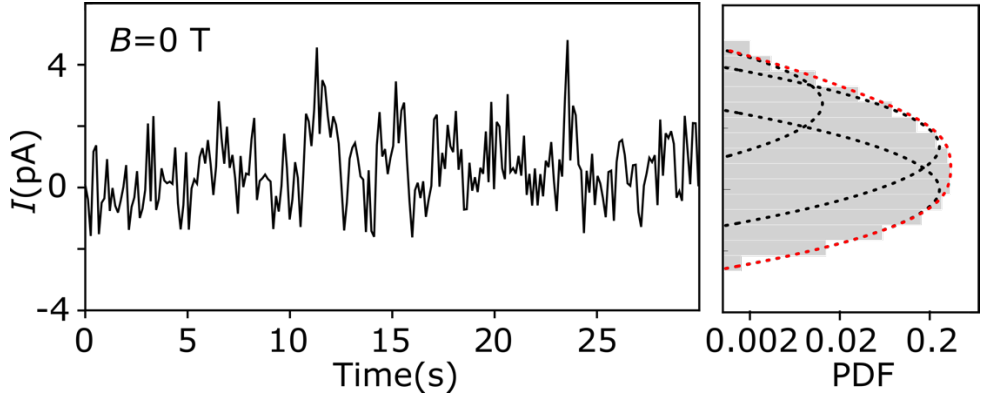

**Figure 4.** Short section of the current trace at the edge of a Coulomb diamond of the **{Co₄}**-functionalized CNT quantum dot and corresponding PDF of the entire signal (1300 s) with three Gaussian peaks fitted at $B = 0$ T.

Studying the derivative of the direct current signal of the **{Mn₄}**-device (see Figure 3b) with respect to time, we extract the largest possible value between two current levels to be $\Delta I \sim 80$ pA and the standard deviation of $\sigma_{deriv} = \sqrt{2}\,\sigma_{direct} = 56.4(1.3)$ pA, with the standard deviation of the direct signal $\sigma_{direct}$ (see Supporting Information section SI-4 for an extended discussion). The PDF of the direct current signal can thus be fitted by equidistant Gaussian peaks with FWHM $= 2\sqrt{2\ln(2)}\,\sigma_{direct} = 94$ pA. This results in at least 14 equidistant current levels. Each level is directly related to an energy that can be evaluated from the current-voltage curve taken at this gate voltage. We find a maximum spacing of $\Delta\varepsilon \sim 27$ μeV. (The level spacing $\Delta\varepsilon$ might be in fact smaller, but this is beyond our resolution. This would lead to more levels, but in fact leave the main results discussed below qualitatively unchanged.)

The PDF of the RTS signal in the {Co₄} functionalized devices decays much faster with increasing current levels than for {Mn₄}. Therefore, only three different current levels with a spacing of $\Delta\varepsilon \sim 50$ μeV are identified.

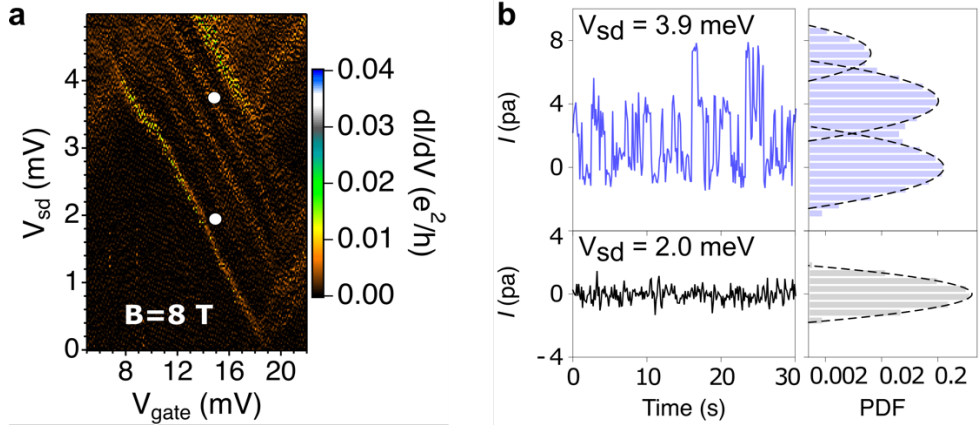

**Figure 5. a**: Part of the stability diagram of a CNT quantum dot functionalized with approx. 70 **{Co₄}** complexes at $B = 8$ T (perpendicular to CNT axis). White dots indicate the positions where the current traces shown in b were measured (gate and bias voltage of the lower point the same as in Fig. 4. **b**: Parts of the time traces of the current taken at the indicated positions in a (top, in blue, for $V_{sd} = 3.9$ mV, bottom, in black, for $V_{sd} = 2.0$ mV) with the respective PDF of the entire direct current traces after background correction on the right. Discrete current levels are indicated by dashed lines.

The larger spin polarization and orbital momentum in the **{Co₄}**-CNT, predicted by DFT, will result in a response to an external magnetic field. Figure 5a presents a part of the stability diagram of the **{Co₄}** device taken at $B = 8$ T. Each line is split into two compared to Figure 2b and corresponds to different Zeeman levels of the CNT quantum dot with $\Delta E_{Zee} = 0.8$ meV. No RTS is observed when $dI/dV = 0$, as the quantum dot serves as detector only for $dI/dV \neq 0$. Exemplary current traces measured at two positions (white dots in Fig. 4a) are shown in Figure 5b. The background noise in the Coulomb blockade and on the plateaus of the Coulomb staircase shows a Gaussian distribution as expected (Figure 5b, lower trace). Between two plateaus where the slope of the current change is steepest the current traces exhibit a clear random telegraph signal (Figure 5b, upper trace).

Indeed, three different current levels with larger energy splitting compared to $B = 0$ T are clearly identified in the respective PDF. We find that the steps between two adjacent energy levels are equidistant with $\Delta\varepsilon \sim 0.1$ meV. Fitting three Gaussian peaks to the PDF we find similar amplitudes for the signals with and without a magnetic field applied.

**Theoretical model**

We interpret the telegraph signal in the following way: from time to time an electron tunneling through the quantum dot provides the energy to excite the combined system of CNT quantum-dot and all attached molecules. The excitation energy $\Delta\varepsilon$ is that of a single {$\mathbf{M_4}$} complex (Figure 6). The excited state is thereafter probed by many following electrons that experience a state-dependent tunneling rate, establishing a characteristic total current. Multiple molecules can be present in an excited state at any given time (with energy $\Delta\varepsilon$ each), leading to the multi-level PDF of the current. For relaxation, the {$\mathbf{M_4}$} complexes dissipate their energy to the quantum-dot electrons.

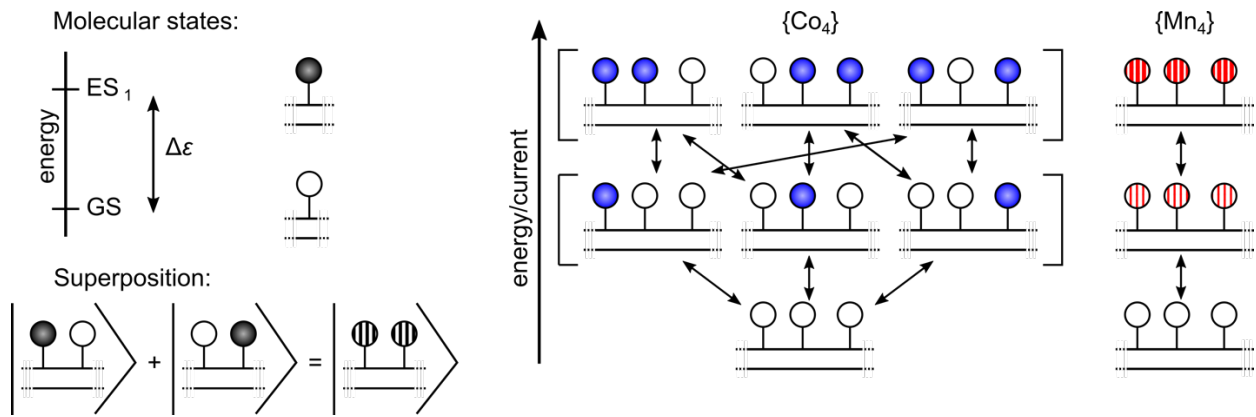

**Figure 6.** Sketch of the different $|S_{\mathrm{tot}}, S_{\mathrm{tot}}^z, S_{12}, S_{34}\rangle$ states of the system. Molecules can be in their ground state (open circles), the first excited state (full circles) or a superposition thereof (striped circles). The higher the current level, the more molecules are in an excited state. Here, only three

of the attached molecules are shown with the possible transitions. In the case of **{Co₄}**, the different transitions are independent. For **{Mn₄}**, the states exist in a superposition and thus, the transitions are coherent.

The energy difference between the excited states sampled by the current in **{Mn₄}** (0.027 meV), and in **{Co₄}** (0.1 meV at $B = 8$ T) is orders of magnitude smaller than other processes known to cause RTS in CNT devices.[36-38] Note, that these transitions do not involve spin flips as those would require comparably large energies, since states with different spin configuration lie about 2 orders of magnitude higher in energy (see SI-2). Instead, different linear superpositions of spin pairs yielding $S_{tot}^z = 0$ differ in energy because of non-equivalent exchange interactions between them (Figure 6 and supporting information section SI-1). Thus, we conclude that the excitation of the quantum dots is due to excitation of non-degenerate molecular $S_{tot} = 0$ eigenstates.

In case of {Co₄}, we see a small dependence of the splitting on the magnetic field. We attribute this to the small residual spin and orbital momentum left on the complex (Tables S3 and S4) as derived from DFT calculations. In total a magnetic moment as small as $\sim 0.4\ \mu_B$ might be left on the complex leading to a Zeeman energy in the of $\sim 100\ \mu eV$. This is an order of magnitude below the antiferromagnetic coupling within the complex.[28] Therefore, the states can be still treated as antiferromagnetic with the magnetic field as small perturbation.

The full time trace of the RTS contains more information than just the relative probability with which the different current levels occur. Since it monitors in time each transition between the different states, the RTS contains information about correlations of the switching events which, in turn, can be used to reveals details about the underlying system. In the present case, we are able to show that there is a fundamental difference in the excitation of the **{Co₄}**/**{Mn₄}** ensembles: For the **{Co₄}**-functionalized CNT, there are *independent* two-level fluctuators that can be excited *individually,* while for the **{Mn₄}**-functionalized CNT the {Mn₄} complexes are *collectively* excited by forming a *coherent* superposition of excitations in *all* molecules.

To reach this conclusion, we analyze the RTS in terms of the full counting statistics. For this, we separate the time trace of the RTS into time intervals of length $\Delta t$ and count the number of transitions from higher to lower levels (or equivalently from lower to higher levels). We, thus, obtain the probability distribution $P_N(\Delta t)$ that $N$ transitions have occurred. This probability distribution can be characterized by so-called factorial cumulants, which are derived in the following way. First, one constructs the generating function $M_F(z, \Delta t) = \sum_{N=0}^{\infty}(z + 1)^N P_N(\Delta t)$, given by the $z$-transform of the probability distribution.[39-42] Performing the $m$-th derivative of the generating function with respect to $z$ leads to the factorial moment of order $m$, which is nothing but the expectation value $\langle N(N - 1) \dots (N - m + 1) \rangle = \partial_z^m M_F(z, \Delta t)|_{z=0}$. The *factorial cumulant* of order $m$ is obtained by the corresponding derivative of the logarithm of the generating function,

$$C_{F,m}(\Delta t) = \langle\langle N(N - 1) \dots (N - m + 1) \rangle\rangle = \partial_z^m \ln M_F(z, \Delta t)|_{z=0}.$$

Factorial cumulants are particularly suited to highlight deviations from a Poisson process since for the latter all factorial cumulants of order $m \geq 2$ vanish. This property makes them also resilient to detector imperfections such as limited time resolution and noise in the measurement apparatus, which may lead to missing or false counts.[43] One very important feature of the factorial cumulants is that, for independent two-level fluctuators, their sign does not change as a function of time and should alternate,[39-42] $C_{F,m}$ (t) $\sim$ (-1)$^{m-1}$, with the order $m$. In particular, for independent two-level fluctuators, the second factorial cumulant has to be *negative*.

This is the case for the **{Co₄}**-functionalized CNT, see blue curves in Figure 7b-d. Assuming 70 **{Co₄}** complexes, we extract from the first factorial cumulant the transition rates. These are used to calculate the second and third factorial cumulant (dashed lines in Figure 7c-d), resulting in good agreement with the measured data at $B = 8$ T. For the **{Mn$_4$}**-functionalized CNT (red curves in Figure 7b-d), the assumption of independent two-level fluctuators can be ruled out due to the wrong sign of the second cumulant. Instead, we assume 50 {Mn$_4$} complexes that are *collectively* excited by forming a *coherent* superposition of excitations in *all* molecules. This model (with transition rates determined from fitting the first cumulant) correctly reproduces the second and third cumulant (Figure 7c-d) as well as the exponential occupation probability distribution of the data (Figure 7a and S7b). The occupation probabilities $P_{occ}$ of the levels depend on the transition rates between adjacent levels (Figure S6). This behavior is incompatible with independent excitations of individual **{Mn$_4$}** complexes as this would yield a binomial distribution of $P_{occ}$. Assuming a thermally activated process, we find an activation temperature of 450 mK, in good agreement with the electron temperature of 600 mK. The signal caused by the {Co$_4$} complexes at $B = 0$ T decays too quickly for a meaningful statistical analysis. However, the occupation probability seems to behave similarly as for $B = 8$ T following a binomial decay rather than an exponential decay, which implies uncorrelated events.

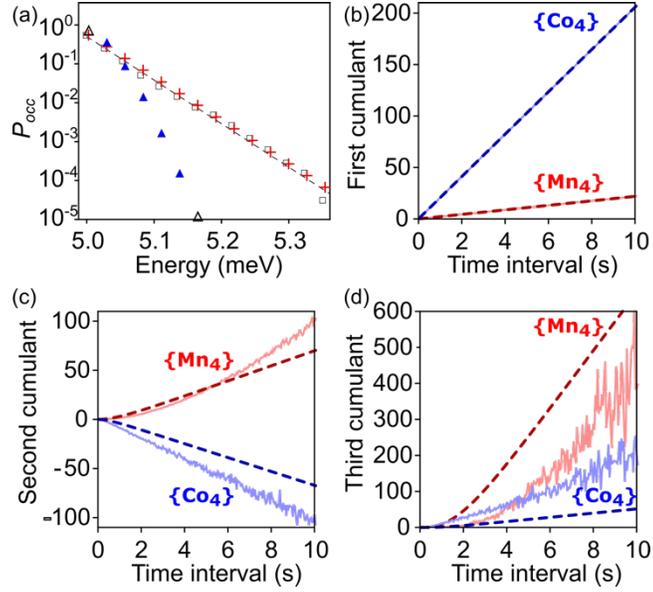

**Figure 7. a**: Occupation probability of the different levels found in the **{Mn₄}**-device on a logarithmic scale for the direct current data (black squares), and for models assuming 50 independent two-level fluctuators (blue triangles), and 50 coherently coupled two-level fluctuators (red crosses). The dashed line is a fit of the data to an exponential distribution. **b**: First factorial cumulant ($C_{F,1}$) as extracted from the measurements on the CNT quantum dots functionalized with **{Mn₄}** (red) **{Co₄}** (blue) complexes. Fits (dashed lines) yield the transition rates used for modelling, in **c**, the second ($C_{F,2}$), and, in **d**, the third ($C_{F,3}$) factorial cumulant. The solid lines in **b**, **c**, **d** are the respective cumulants extracted directly from the digitized measurement signal of the two devices ({Mn₄}: red, {Co₄}: blue). The positive second cumulant in **c** is related to a super-Poissonian Fano factor by $F = (C_{F,2}/C_{F,1}) + 1$. All lines of the **{Co₄}** sample are multiplied by a factor of 20 for better visibility.

Since a model of independent three-level fluctuators also fails to reproduce the data (Figure S7b), we conclude that long-lived coherent excitations involving *all* **{Mn₄}** complexes are present in our system. The coherence is due to the fact that the excitation is caused by one single electron that is coherent in the CNT quantum dot and therefore couples to all molecules at the same time. The long coherence time in the order of seconds is, indeed, likely if both the ground and the excited state of the {**Mn₄**} complexes do not carry a magnetic moment that could couple to any dipolar (magnetic or electric) field.

In the {**Co₄**} device, on the other hand, the $Co^{II}$ ions in their octahedral ligand fields possess a considerable orbital moment. Though the applied magnetic field is not large enough to change the antiferromagnetic nature of the ground state (SI-2), the residual orbital moment couples to fluctuations of dipolar fields leading to fast decoherence due to spin-orbit coupling.[44,45]

**Conclusion**

In the present work, we study the random telegraph signal (RTS) of carbon nanotube quantum dots functionalized with small ensembles of two molecular analogues of antiferromagnets (AFMs), {**Mn₄**} and {**Co₄**}. While the energy changes corresponding to the transitions between different current levels are very similar for both complexes, the statistical behavior of the RTS differs fundamentally between the manganese and cobalt derivatives. As the only fundamental difference between the two complexes concerns the orbital magnetic moment, which is fully quenched in {**Mn₄**} but present in {**Co₄**}, the RTS is necessarily linked to magnetic transitions, specifically transitions between non-degenerate $S_{tot} = 0$ states of individual molecular antiferromagnets. Based on the RTS statistics we can show that while complexes based on magnetic $Co^{II}$ ions switch independently, their congeners based on $Mn^{II}$ ions exhibit a long-lived coherent superposition between the states of all molecules attached to the quantum dot. This leads to a fundamentally new perspective on information processing with magnetic molecules. Molecular antiferromagnetic coupling between individual molecules in the order of 100 K is well known.[46] Hence, it appears feasible to design antiferromagnetic molecules with non-degenerate $S_{tot} = 0$ eigenstates that exhibit similar strong coupling, causing a larger separation in energy than the molecules presented here, enough to prevent switching between states caused by temperature. Instead, by inducing dipolar moments, Raman transitions could be driven to change states deliberately. Alternatively, similar control could be reached using tunable exchange interactions $J$ to modify the energetic order of $S_{tot} = 0$ states. The difference in behaviour of both molecular systems investigated here reveals the large impact of even small residual orbital moments on the spin coherence in molecular AFMs. In the complete absence of such moments strikingly long coherence times in the order of seconds can be achieved. The coupling of the molecules to each other is mediated exclusively by the small but long-ranged spin polarization transferred to the conduction electrons of the CNT. This offers means of controlling the device: using ferromagnetic contacts, spin-polarized electrons can be injected into the CNT[47] and the spin polarization of the conduction electrons can be controlled using a gate.[48]

ACKNOWLEDGMENT

The authors thank Christopher Nakamoto for helping with carrying out the synthesis and crystallisation of the tetranuclear complexes and Christian Lurz for valuable discussions about the data analysis. We acknowledge financial support by the Deutsche Forschungsgemeinschaft (DFG) under Project-ID 278162697 – SFB 1242 as well as for individual grants numbers MA 4079/10-1, ME 3275/6-1, and ZA 780/3-1. PS acknowledges support from the German National Academy of Sciences Leopoldina (Grant No. LPDS 2019-10). Furthermore, ZZ acknowledges funding by the Ramón y Cajal programme RYC-2016-19344 (MINECO/AEI/FSE, UE), the Netherlands sectorplan program 2019-2023, and the research program "Materials for the Quantum Age" (QuMat, registration number 024.005.006), part of the Gravitation program of the Dutch Ministry of Education, Culture and Science (OCW). ZZ, SA and NW acknowledge computer time from PRACE on Archer (EU grant RI-653838) and on MareNostrum4 at Barcelona Supercomputing

Center (BSC), Spain (OptoSpin project id. 2020225411), from JARA-HPC (project JHPC39), and from RES (projects FI-2020-1-0014, FI-2020-1-0018, FI-2020-2-0034) on MareNostrum4. ZZ, SA, NW and PO acknowledge the EC H2020-INFRAEDI-2018-2020 MaX Materials Design at the Exascale CoE (grant No. 824143), Grant PGC2018-096955-B-C43 funded by Spain's MCIN / AEI /10.13039/501100011033 and by ERDF - A way of making Europe, Spain's Severo Ochoa Centers of Excellence Program (Grant No.SEV-2017-0706), and Generalitat de Catalunya CERCA programme (No. 2021 SGR 00997). NW acknowledges funding from the EU-H2020 research and innovation programme under the Marie Sklodowska-Curie programme (Grant No. 754558). This work benefited from the access provided by ICN2 (Barcelona, Spain) within the framework of the NFFA-Europe Transnational Access Activity (grant agreement No 654360, proposal ID 753, submitted by CM). J.M. and A.V. acknowledge funding under ERC grant no. 259286. C.M. acknowledges funding by Niedersächsisches Vorab Akz. 11-76251-14-3/15(ZN3141).

**Appendices**

**Appendix A: Atomistic simulations**

First principles simulations of the {$M_4$}-CNT systems (with M = Mn, Co) were performed using the SIESTA[49] implementation of Density Functional Theory (DFT), within the Local Spin Density Approximation (LSDA). The {$M_4$} complex are bound via a carboxylate (-$CO_2^-$) group to the dangling C of a mono-vacancy site on the external wall of a metallic armchair (5, 5) nanotube. The choice of a metallic nanotube ensures that electron conduction is allowed around the Fermi energy. We have shown that the oxygenated vacancy is especially favorable for functionalization of CNTs with $CO_2$[29] or magnetic nanoparticles.[30] We performed simulations of a single molecule grafted to an infinite tube using the *open-system* set up of Ref. [31], combining the non-equilibrium Green's function formalism with a DFT Hamiltonian as implemented in TranSIESTA.[50] This is necessary to avoid artefacts resulting from the long range character of the indirect exchange coupling between magnetic clusters mediated by the conduction electrons of the carbon nanotube.[51] The simulation set-up was composed of a central region of 35 CNT rings (86 Å long) contacted by two semi-infinite (5,5)-CNTs on each side. The spin-orbit coupling was introduced using the formalism of Ref. [52], as implemented in SIESTA [53]. Further technical details are provided in supporting information section SI-2.

**Appendix B: Device fabrication**

Electronic devices were fabricated as described in Schnee et al.[32] Briefly, isolated single wall carbon nanotubes were grown on a $Si/SiO_2$ substrate using methane as a feedstock and Fe/Mo catalyst nanoparticles. The nanotubes were oxidized in air at 450° and then the field effect transistor structure of the quantum dot device was fabricated using the substrate as back gate and Pt leads (**{Mn$_4$}**-based devices) or the device was fabricated and then the nanotubes oxidized (**{Co$_4$}**-based devices). $[M_4L_2(OAc)_4]$ complexes ($H_2L = 2,6$-bis-(1-(2-hydroxyphenyl) iminoethyl)pyridine, **{Mn$_4$}** and **{Co$_4$}**) were synthesized according to the literature,[26,27] and bonded to the nanotubes by introducing the device in a solution of the complex (in acetonitrile solution for **{Mn$_4$}**, in dichloromethane for **{Co$_4$}**) for a week before rinsing away the non-covalently attached complexes over another 1 week. A density of functionalization of ca. 1 molecule/10 nm was achieved.[28]

**Appendix C: Electronic measurements**

All electronic measurements were carried out in a $^3He/^4He$ dilution refrigerator at a base temperature around 100 mK. The bias voltage was supplied symmetrically via digital-to-analogue converters, addressed via an optical fiber in order to decouple it from the mains. The electron temperature was determined from the width of the coulomb-peaks and independently by fitting a Fermi-function to the first current step (ground state tunneling) of the Coulomb-staircase taken from the {Co₄}-CNT stability diagram. The current of the noise traces was amplified using a low-noise current-to-voltage (IV) converter (QT-IVVI Rack, TU-Delft) and was measured by averaging over 20 ms (NPLC 1) using an HP34401A multimeter. In total we had four devices (one based on {Mn₄}, three based on {Co₄}) that exhibited regular Coulomb diamonds after cool-down. All of them exhibited random telegraph signals. Two of the devices are discussed in this paper.

## Appendix D: Raman spectroscopy

Confocal micro-Raman spectroscopy was performed at room temperature in backscattering geometry with a Horiba Jobin Yvon LabRAM HR800 monochromator. For this, a He-Ne laser emitting at 633 nm was focused through a 100x microscope objective with 0.95 NA, and the power was kept at 0.7 mW (as measured before the objective). With the help of previously acquired SEM images of the device,[32] the sample was oriented such that the nanotube is aligned parallel to the laser polarization in order to maximize the signal. The target area was mapped with a motorized stage to find the precise nanotube's position through its Raman signal, and the depth-focus was subsequently adjusted on the nanotube for maximal intensity.

The Raman spectra were spectrally calibrated by the emission lines of a neon lamp. The background signal that is caused by Raman scattering with the device's silicon wafer was measured with ~2 μm lateral distance to the nanotube, which allowed for subsequent subtraction from the signal.

## Appendix E: Data analysis

We used the histogram of the derivative of the direct current data (before background correction) of the {Mn₄} functionalized device to determine current steps of $\Delta I \approx 80$ pA with a standard deviation of $\sigma_{direct} = 40(1)$ pA (see **Figure 3b**). To digitize the random telegraph signal, the background of the noise traces was removed using the statistics-sensitive nonlinear iterative peak clipping (SNIP) algorithm implemented in the computer algebra system Wolfram *Mathematica*. The PDFs and cumulants were determined from a current time trace of 1300 s (3600 s) length for the **{Co$_4$}**-functionalized (**{Mn$_4$}**-functionalized) device. The models used to simulate the data are elaborated in the supporting information.

ASSOCIATED CONTENT

**Supporting Information Available:** Discussion of spin states in the **{M$_4$}** complex, DFT calculations of **{Mn$_4$}** or **{Co$_4$}** complex grafted to the carbon nanotube, Raman spectrum of the **{Mn$_4$}** functionalized CNT device, histogram of the derivative of the original data, theoretical model and results (PDF file); coordinates of CNT-**{M$_4$}** systems used for DFT calculations (ZIP archive containing separate XYZ files).

**Supplemental Information**

**Independent and coherent transitions between antiferromagnetic states of few-molecule systems**

Claire Besson, Philipp Stegmann, Michael Schnee, Zeila Zanolli, Simona Achilli, Nils Wittemeier, Asmus Vierck, Robert Frielinghaus, Paul Kögerler, Janina Maultzsch, Pablo Ordejón, Claus M. Schneider, Alfred Hucht, Jürgen König, Carola Meyer*

*Correspondence should be addressed to carola.meyer@uos.de

**SI-1 Discussion of spin states in the {M₄} complex**

In our effective spin model, we use the fact that each of the four $Mn^{II}$ ions, $i = 1, \cdots, 4$ carries a spin $\vec{S}_i$ with spin quantum number $S_i = \frac{5}{2}$, and magnetic quantum number $S_i^z = -\frac{5}{2}, -\frac{3}{2}, \cdots, \frac{5}{2}$. This yields a $6^4$-dimensional Hilbert space for the spin states. Instead of characterizing the spin states by the quantum numbers $\{S_1^z, S_2^z, S_3^z, S_4^z\}$, we can use the standard procedure for adding angular momenta to find a basis that is more convenient for our purpose. Let $S_{ij}$ be the spin quantum number and $S_{ij}^z$ be the magnetic quantum number of the spin sum $\vec{S}_{ij} := \vec{S}_i + \vec{S}_j$. In a first step, we combine spins 1 with 2 and 3 with 4 (see Fig. S1). This leads to a new basis characterised by $\{S_{12}, S_{12}^z, S_{34}, S_{34}^z\}$. In a second step, we form the total spin

$$\vec{S}_{\text{tot}} := \sum_{i=1}^{4} \vec{S}_i = \vec{S}_{12} + \vec{S}_{34} \tag{1}$$

with the spin quantum number $S_{\text{tot}}$ and the magnetic quantum number $S_{\text{tot}}^z$. The resulting basis can be labelled by the set of quantum numbers $\{S_{\text{tot}}, S_{\text{tot}}^z, S_{12}, S_{34}\}$. (Alternatively, we could have chosen $\{S_{\text{tot}}, S_{\text{tot}}^z, S_{13}, S_{24}\}$ or $\{S_{\text{tot}}, S_{\text{tot}}^z, S_{14}, S_{23}\}$.)

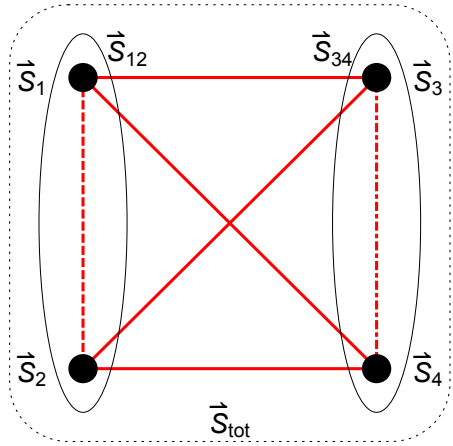

**Figure S1 | Schematic view of the spin system of the {Mn₄} complex.** Four $\frac{5}{2}$ - spins are coupled to each other. If all exchange interactions are equal, the ground state is six-fold degenerate. For small deviations between the interactions, the degeneracy is split into six antiferromagnetic eigenstates of different energy.

We assume a Heisenberg-like spin-exchange coupling for each pair of Mn spins,

$$H = -\frac{1}{2} \sum_{i \neq j} J_{ij} \vec{S}_i \cdot \vec{S}_j \tag{2}$$

with antiferromagnetic $J_{ij} < 0$.

If all exchange couplings are equal, $J_{ij} = J$, we can use $\sum_{i \neq j} \vec{S}_i \cdot \vec{S}_j = \vec{S}_{\text{tot}}^2 - \sum_i \vec{S}_i^2$, together with $\vec{S}_{\text{tot}}^2 = S_{\text{tot}}(S_{\text{tot}} + 1)$ and $\vec{S}_i^2 = S_i(S_i + 1) = \frac{35}{4}$ to see that the eigenenergy

$$E = -\frac{J}{2}[S_{\text{tot}}(S_{\text{tot}} + 1) - 35] \quad (3)$$

only depends on the spin quantum number $S_{\text{tot}}$ of the total spin but not on $S_{\text{tot}}^z$, $S_{12}$, or $S_{34}$. For antiferromagnetic coupling, $J < 0$, the ground state has total spin $S_{\text{tot}} = 0$ with vanishing dipole moment, $S_{\text{tot}}^z = 0$. It is six-fold degenerate with

$$|S_{\text{tot}}, S_{\text{tot}}^z, S_{12}, S_{34}\rangle = |0,0,0,0\rangle, |0,0,1,1\rangle, |0,0,2,2\rangle, |0,0,3,3\rangle, |0,0,4,4\rangle, |0,0,5,5\rangle$$

Unequal $J_{ij}$'s remove this degeneracy. Instead of a six-fold degenerate ground state, there are now six energy eigenstates, all with $S_{\text{tot}} = S_{\text{tot}}^z = 0$, as long as the variation between the different $J_{ij}$'s is not too large. States with $S_{\text{tot}} > 0$ remain higher in energy. Therefore, we can conclude that both the ground and the first excited state do not carry any dipole moment, i.e. $S_{\text{tot}}^z = 0$.

As an example, assuming $J_{12} < J_{34} < J_{31} = J_{14} = J_{23} = J_{24}$ the ground state $|\Psi_0\rangle = |0,0,0,0\rangle$ and the first excited state $|\Psi_1\rangle = |0,0,1,1\rangle$ (in the $S^z$-base) are given by:

Ground state:

$|\Psi_0\rangle = |\varphi_0\rangle \otimes |\varphi_0\rangle$ with

$$|\varphi_0\rangle = \frac{1}{\sqrt{6}} \left( \left|\frac{5}{2}, -\frac{5}{2}\right\rangle - \left|\frac{3}{2}, -\frac{3}{2}\right\rangle + \left|\frac{1}{2}, -\frac{1}{2}\right\rangle - \left|-\frac{1}{2}, \frac{1}{2}\right\rangle + \left|-\frac{3}{2}, \frac{3}{2}\right\rangle - \left|-\frac{5}{2}, \frac{5}{2}\right\rangle \right).$$

First excited state:

$|\Psi_1\rangle = a_{-1}|\varphi_{1,-1}\rangle \otimes |\varphi_{1,1}\rangle - a_0|\varphi_{1,0}\rangle \otimes |\varphi_{1,0}\rangle + a_1|\varphi_{1,1}\rangle \otimes |\varphi_{1,-1}\rangle$ with

$$|\varphi_{1,1}\rangle = \frac{\sqrt{10}\left|+\frac{5}{2}, -\frac{3}{2}\right\rangle - 4\left|+\frac{3}{2}, -\frac{1}{2}\right\rangle + 3\sqrt{2}\left|+\frac{1}{2}, +\frac{1}{2}\right\rangle - 4\left|-\frac{1}{2}, +\frac{3}{2}\right\rangle + \sqrt{10}\left|-\frac{3}{2}, +\frac{5}{2}\right\rangle}{\sqrt{70}},$$

$$|\varphi_{1,0}\rangle = \frac{5\left|+\frac{5}{2}, -\frac{5}{2}\right\rangle - 3\left|+\frac{3}{2}, -\frac{3}{2}\right\rangle + \left|+\frac{1}{2}, -\frac{1}{2}\right\rangle + \left|-\frac{1}{2}, +\frac{1}{2}\right\rangle - 3\left|-\frac{3}{2}, +\frac{3}{2}\right\rangle + 5\left|-\frac{5}{2}, +\frac{5}{2}\right\rangle}{\sqrt{70}},$$

$$|\varphi_{1,-1}\rangle = \frac{\sqrt{10}\left|+\frac{3}{2}, -\frac{5}{2}\right\rangle - 4\left|+\frac{1}{2}, -\frac{3}{2}\right\rangle + 3\sqrt{2}\left|-\frac{1}{2}, -\frac{1}{2}\right\rangle - 4\left|-\frac{3}{2}, +\frac{1}{2}\right\rangle + \sqrt{10}\left|-\frac{5}{2}, +\frac{3}{2}\right\rangle}{\sqrt{70}}.$$

The coefficients verify $a_i = \frac{1}{\sqrt{3}}$ for magnetic field $B = 0$. They differ in case a magnetic field is applied.

The same procedure applies to M = Co, except for the difference in spin quantum number, which is only 3/2 for Co$^{\text{II}}$, so that the singlet ground state is only 4-fold (instead of 6-fold) degenerate in the case of a perfect magnetic tetrahedron. Again, the degeneracy between those states is lifted as the $J$ couplings between centers are not identical in the actual complex.

**SI-2 DFT calculations of {Mn₄} and {Co₄} complex grafted to the carbon nanotube.**

The structural, electronic and magnetic properties of the **{M₄}**-CNT hybrid systems (where M = Mn or Co) are computed from first principles using the SIESTA[1-2] code. The **{M₄}** complex is bound via a -$CO_2^-$ group to the dangling C of a mono-vacancy site on the external wall of a metallic armchair (5, 5) nanotube. The choice of a metallic nanotube ensures that electron conduction is allowed around the Fermi energy. Previous studies have demonstrated that the mono-vacancy site is especially favorable for functionalization of CNTs with molecules[3] or magnetic nanoparticles.[4]

In order to take into account the strong correlation of the 3d electrons and avoid the excessive delocalisation of the *d* states predicted in the Local Density Approximation, Hubbard-like corrections $U = 6$ eV and $U = 4$ eV were used for Mn and Co, respectively (LDA+U method).[5] The same $U = 6$ eV value was used by Kampert *et al.* in their calculations on **{Mn₄}**.[6] A standard double zeta polarised (DZP) basis set was used for carbon, nitrogen and hydrogen, and an optimised double-zeta (DZ) for Mn, Co and O. Calculations were spin polarised and performed assuming collinear spins. LDA+SOC calculations (off-site formalism of Ref. [7]) without Hubbard correction were performed in order to determine the orbital moments and the role of SOC as currently, the SIESTA code does not allow to include both SOC and Hubbard correction. We verified that the effect of spin-orbit interaction is negligible in **{Mn₄}** (as expected for a half filled *3d* shell) but not in **{Co₄}**. Convergence of electronic structure and magnetic properties was achieved for a real space grid cut-off of 400 Ry, and a Fermi-Dirac smearing of 100 K in the LDA+U calculation, while with SOC a cut-off of 650 Ry and electronic temperature of 1 K were adopted. The atomic positions were relaxed in standard periodic boundary conditions simulations, with a $1 \times 1 \times 12$ k-points sampling of the Brillouin zone for 15 cells of **{M₄}**-CNT (shifted grid), and the conjugate gradient algorithm. The simulation cell extends for 36.9354 Å (30 carbon atoms) along the periodic direction while more than 30 Å of vacuum between periodic replica of the system have been taken in the two directions perpendicular to the tube axis. The maximum force on atoms was smaller than 0.04 eV/Å for the CNT+**{M₄}** system. Open-system simulations were performed within a non-equilibrium Green's function formalism, using the TranSIESTA solution method,[8-9] on a 70-carbons long units consisting of the relaxed {M₄-CNT} unit padded with (5,5)-CNT fragments (20 carbons long total) on either side.

## 1) Periodic boundary condition simulations

### a) {M₄}-CNT ground state magnetic configuration

Collinear-spin PBC simulations were performed for **{M₄}**-CNT systems differing for the direction of the spin magnetic moment of the four magnetic atoms in the complex: aligned parallel (up) or antiparallel (down) to an arbitrary direction. We find that the spin configuration with lowest total energy (ground state) has total spin $S_{tot}^z = 0$ for both isolated[10] and grafted molecules, and for both complexes. The ground state is labelled up-up-down-down (*uudd*) referring to the relative alignment of the four M ions. The M ions connected to the same kind of ligand are almost equivalent, as they face a similar chemical environment. In our notation, the pairs of quasi-equivalent ions are M1/M2 and M3/M4. The ground state presents an antiferromagnetic (AFM) coupling between non-equivalent M ions (M1/M3, and M2/M4), and a ferromagnetic (FM) coupling between the equivalent pairs (M1/M2 and M3/M4).

The total energy of the investigated spin configurations for the fully relaxed **{Mn₄}**-CNT and **{Co₄}**-CNT are reported in Tables S1 and S2, together with the magnetization of the four M atoms. As in the free-standing case,[10] the absolute values of the magnetic moments of the two not equivalent pairs of magnetic atoms are slightly different. However, the molecules maintain a total spin $S_{tot}^z \sim 0$ in the AFM configurations, due to the spin polarization of the ligands, which is more relevant in **{Co₄}**.

| Config. | $\Delta E$ [meV] | $\mu^{S}{}_1$ [$\mu_B$] | $\mu^{S}{}_2$ [$\mu_B$] | $\mu^{S}{}_3$ [$\mu_B$] | $\mu^{S}{}_4$ [$\mu_B$] |
|---------|---------|---------|---------|---------|---------|
| **uudd** | **0.0** | **4.82** | **4.82** | **-4.88** | **-4.87** |
| uddu | 2.2 | 4.82 | -4.82 | -4.88 | 4.87 |
| uduu | 3.7 | 4.82 | -4.82 | 4.88 | 4.87 |
| uuud | 5.1 | 4.82 | 4.82 | 4.88 | -4.87 |
| uddd | 5.5 | 4.82 | -4.82 | -4.88 | -4.87 |
| uuuu | 8.0 | 4.83 | 4.82 | 4.88 | 4.87 |
| udud | 12.8 | 4.82 | -4.82 | 4.88 | -4.87 |

**Table S1 |** Low-energy magnetic configurations for the **{Mn₄}**-CNT obtained from periodic boundary conditions DFT+U calculations. The ground state (uudd) is highlighted in bold. We report the total energy with respect to the ground state configuration ($\Delta E$, in meV), and the spin magnetic moment for each ion. The units of the magnetic moment are Bohr magnetons ($\mu_B$). The data refer to the 15 cells PBC calculation.

| Config. | $\Delta E$ [meV] | $\mu^{S}{}_1$ [$\mu_B$] | $\mu^{S}{}_2$ [$\mu_B$] | $\mu^{S}{}_3$ [$\mu_B$] | $\mu^{S}{}_4$ [$\mu_B$] |
|---------|---------|---------|---------|---------|---------|
| **uudd** | **0.0** | **2.70** | **2.71** | **-2.72** | **-2.73** |
| dddu | 10.0 | -2.71 | -2.71 | 2.72 | -2.74 |
| udud | 10.3 | -2.70 | 2.71 | -2.72 | 2.74 |
| uudu | 10.9 | 2.70 | 2.71 | 2.72 | -2.73 |
| uddd | 14.9 | -2.71 | 2.71 | -2.72 | -2.74 |
| uuuu | 18.1 | 2.71 | 2.71 | 2.73 | 2.74 |
| uddu | 20.3 | -2.65 | 2.79 | 2.65 | -2.79 |

**Table S2 |** Low-energy magnetic configurations for the **{Co₄}**-CNT obtained from periodic boundary conditions DFT+U calculations. The ground state (uudd) is highlighted in bold. We report the total energy with respect to the ground state configuration ($\Delta E$, in meV), and the spin

magnetic moment for each magnetic ion. The units of the magnetic moment are Bohr magnetons ($\mu_B$). The data refer to the 15 cells PBC calculation.

### b) {M₄}- CNT electronic band structure

All the hybrid **{M₄}**-CNT-systems we modelled are metallic, as illustrated by their electronic band structure (Fig. S2). The **{M₄}** complex does not perturb much the electronic structure of the nanotube in the vicinity of the Fermi energy, indicating that the interaction between nanotube and molecule is weak and explaining why the magnetic properties of the **{M₄}** complex are preserved after bonding to the nanotube. The weak interaction between tube and molecule only causes minor changes in the band structure of the other explored magnetic configurations (not shown). The band structure of the two systems looks very similar, featuring the dispersing bands of CNT (5,5) that are slightly spin polarised and dispersionless (flat) bands whose spin polarisation is larger. These bands that lie above the Fermi level (at ~0.5 eV for **{Mn₄}** and ~0.3 eV for **{Co₄}** in the LDA+U calculation – on the left) correspond to energy levels localised on the **{M₄}** complex. In the calculation with SOC (right) the bands show the same properties apart for a shift of the dispersionless bands toward the Fermi level (for both occupied and empty states) due to the absence of the Hubbard U correction that accounts for the electronic correlation.

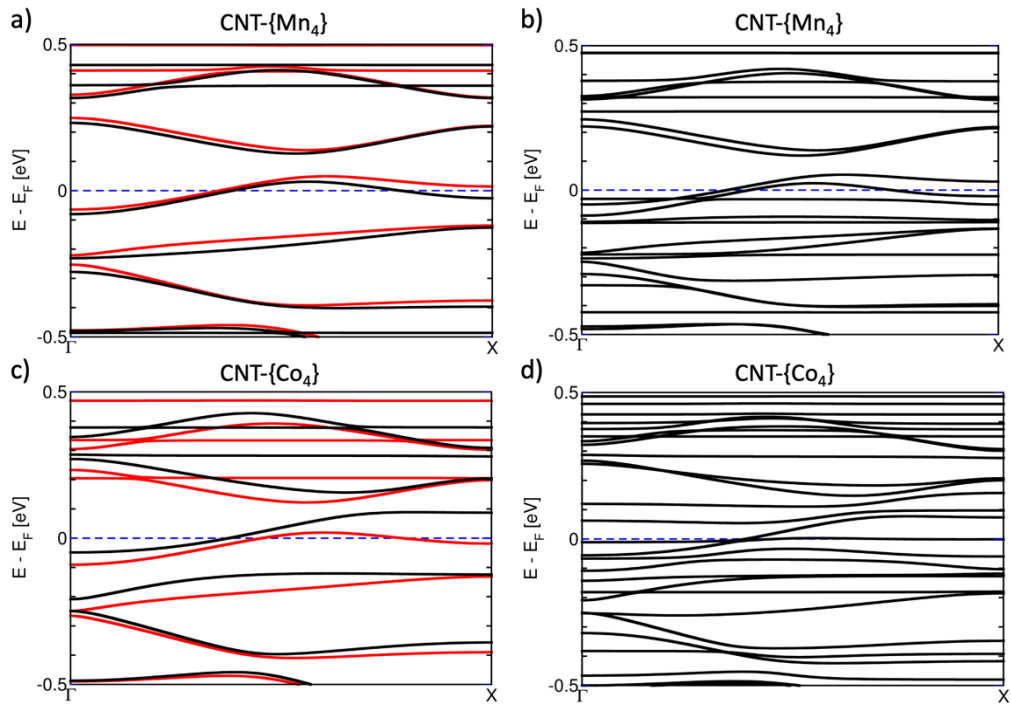

**Figure S2 |** Electronic band structure of a CNT functionalised with a **{Mn₄}** or **{Co₄}** complex computed in the LDA+U approximation (red: majority spins, black: minority spins, panels a and c), or explicitly including SOC (panel b and d). Flat bands in the electronic structure correspond to the energy levels of the complex. The Fermi level is indicated with a dashed blue line.

## 2) Open system simulations

### a) LDA+U: charge transfer between molecule and tube and localized magnetic moments

Charge transfer was computed in the open system set up. Regardless of the spin configuration, the **{Mn₄}** complex always *withdraws* electron charge from the CNT. In the ground state *uudd* configuration the dangling C in the CNT donates 0.08 electrons while the electronic charge transferred from the other eight C atoms near the -$CO_2^-$ group (highlighted in yellow in Fig. S3) is 0.1 electrons, for a total of 0.18 electrons donated by the atoms surrounding the monovacancy.

The charge transfer is different for majority and minority spin electronic charge leading to a small magnetic moment of the molecule ($\mu_{\{Mn4\}}$= -0.006 $\mu_B$) in the ground state configuration, differently from the free-standing case where it was exactly zero. An induced spin polarization is found on the CNT giving $\mu_{\{CNT\}}$= -0.025 $\mu_B$ in the cell. The associated spin density is delocalised over the whole CNT as shown in Fig. 1 of the main manuscript. The C acting as linking site retains a magnetic moment of -0.021 $\mu_B$ while the other eight C atoms belonging to the defect site have a total magnetic moment of -0.004 $\mu_B$. Even though the induced spin polarisation is most important at the functionalized vacancy site, the effect is long ranged and allows for a correlation among the spin state of various molecules grafted to the tube.

In **{Co₄}**-CNT ground state the charge transfer from the linking C to the molecule amounts to 0.13 electrons, larger than in **{Mn₄}**-CNT while for the other eight C atoms near the -$CO_2^-$ group it is 0.03 electrons, i.e. smaller than in the other system. The linking C has an induced magnetic moment of 0.084 $\mu_B$ in **{Co₄}**-CNT which is larger, in modulus, than the value found in **{Mn₄}**-CNT. On the other hand, the sum of the magnetic moments of the other eight atoms forming the defect in the CNT is 0.006 $\mu_B$ in **{Co₄}**-CNT. The magnetic moment on the molecule is 0.021 $\mu_B$ while the net spin polarization induced on the portion of the CNT in the unit cell is 0.117 $\mu_B$ i.e. larger than the other compound ($\mu_{\{CNT\}}$ = −0.03 $\mu_B$).

The percentage of the magnetic moment of the C atoms of the CNT in the cell which is localized on the C atoms of the defect at the linking site (linking C + 8 C atoms) is 76% for **{Co₄}**-CNT and 95% for **{Mn₄}**-CNT.

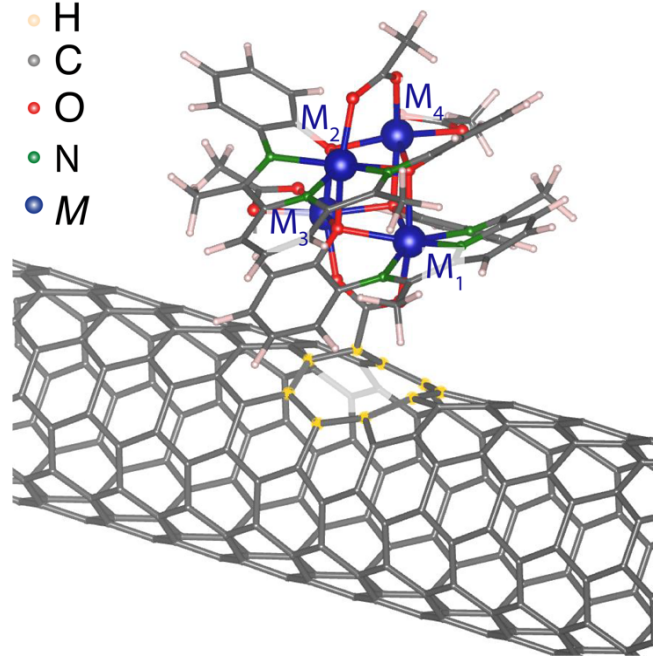

**Figure S3** | Linking of the molecule to the CNT. Color code: C - grey, O - red, N - green, Mn/Co - blue. The atoms surrounding the mono-vacancy site where the molecule is grafted to the tube are highlighted in yellow.

### b) SOC: Orbital moment on the CNT-{**M₄**} hybrid systems

The magnetic moments and orbital angular moments on the individual magnetic sites computed for the open system with SOC and without Hubbard correction are given Table S3. The data refer to the ground-state (uudd) configuration. We note a slight reduction of the magnetic moments compared to the calculation with Hubbard correction and without SOC (Table S4): approximately 0.25 $\mu_B$ for {**Co₄**}-CNT and 0.15 $\mu_B$ for {**Mn₄**}-CNT. The orbital angular moments are finite in {**Co₄**} due to the 3d$^7$ atomic configuration and are almost negligible in {**Mn₄**} due to the half filling of the 3d shell.

| SOC | $\mu^{S_1}$ [$\mu_B$] | $\mu^{S_2}$ [$\mu_B$] | $\mu^{S_3}$ [$\mu_B$] | $\mu^{S_4}$ [$\mu_B$] | $\mu^{L_1}$ [$\mu_B$] | $\mu^{L_2}$ [$\mu_B$] | $\mu^{L_3}$ [$\mu_B$] | $\mu^{L_4}$ [$\mu_B$] |
|---|---|---|---|---|---|---|---|---|
| {**Mn₄**} | 4.55 | 4.55 | -4.65 | -4.61 | 0.06 | 0.06 | 0.07 | 0.07 |
| {**Co₄**} | 2.56 | 2.54 | -2.60 | -2.55 | 0.16 | 0.14 | 0.18 | 0.25 |

**Table S3** | Magnetic moments and orbital angular moments on the individual magnetic sites of {**Mn₄**}-CNT and {**Co₄**}-CNT computed from first-principles in the open-system set-up with SOC.

| LDA+U | $\mu^{S_1}$ [$\mu_B$] | $\mu^{S_2}$ [$\mu_B$] | $\mu^{S_3}$ [$\mu_B$] | $\mu^{S_4}$ [$\mu_B$] | $\mu^{S}_{\text{Molecule}}$ |
|---|---|---|---|---|---|
| {**Mn₄** } | 4.82 | 4.82 | -4.88 | -4.87 | -0.031 |
| {**Co₄**} | 2.71 | 2.70 | -2.73 | -2.72 | 0.138 |

**Table S4** | Magnetic moments on the individual magnetic sites in {**Mn₄**}-CNT and {**Co₄**}-CNT and residual magnetic moment of the whole {**Mn₄**} and {**Co₄**} molecules computed from first-principles in the open-system set-up with LDA+U.

## SI-3 Raman spectrum of the {Mn$_4$} functionalised CNT device

The Raman spectrum shows a metallic tube according to the strong broadening of the $G^{(-)}$-mode[11]; resonance behaviour and position of $G^{(-)}$ and of the TO+ZA combination mode[12] allow for an assignment to the family $2n + m = 30$ (where $(n, m)$ is the CNT chiral index) and the absence of the TO peak suggests a zigzag (15,0) CNT, where the TO mode is Raman-inactive.

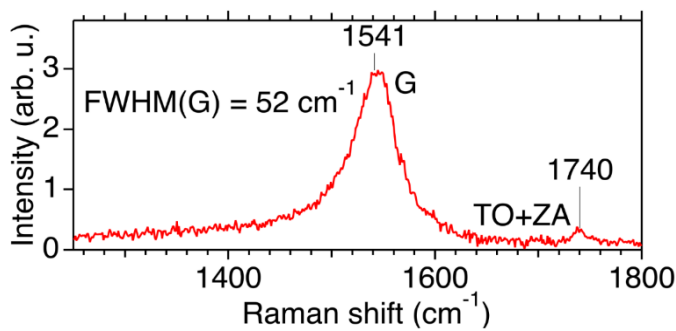

**Figure S4 | Raman spectrum of the {Mn$_4$} functionalised CNT device.** Shape and position of the $G^{(-)}$-mode and the TO+ZA combination mode in combination with the absence of the TO peak suggest a (15,0) zigzag-CNT.[13]

**SI-4 Histogram of the derivative of the direct current data**

Since the influence of the slow fluctuations (time scale 10 seconds) on the derivative with (almost) equidistant points with $\Delta t \approx 100$ ms is negligibly small, we use the derivative of the original data to determine the expected standard deviation of the direct current data $\sigma_{direct}$ which is directly related to the standard deviation of the derivative $\sigma_{deriv}$ with $\sigma_{direct} = \sigma_{deriv}/\sqrt{2}$.

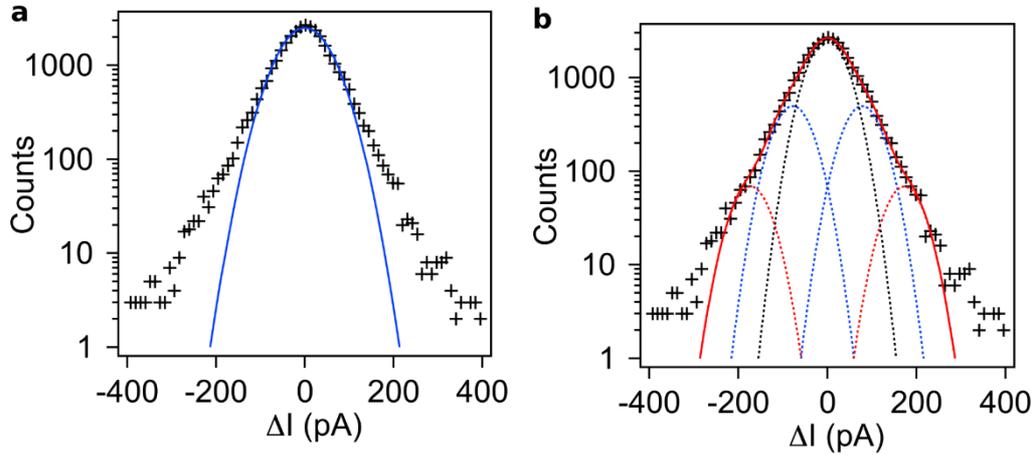

**Figure S5** | Histogram of the derivative of the current time traces (black crosses). **a** Fit with a single gaussian peak (blue solid line) **b** Fit with five Gaussian peaks symmetric around 0 (red solid line) revealing the transitions for one current level (blue dotted line) and two current levels (red dotted lines).

Figure S5 compares the fit of the histogram with a single gauss peak with the fit of the histogram with five Gaussian peaks. It is obvious that the data (black crosses) cannot be described with a single gauss peak fit (Fig. S5a, blue solid line). Additional peaks are necessarily symmetric around zero, since the system is clearly neither continuously excited nor relaxed below a ground level. We assume the same standard deviation for all current levels and thus all transitions between levels. Assuming five Gaussian peaks with the second pair of Gaussian peaks (Fig S5b, red dotted lines) have twice the $\Delta I$ of the first pair (Fig S5b, blue dotted lines) yields the best fit. The latter constrain does not change the residuum compared to a free position as additional fit parameter. We thus end up with a set of five fit parameters for the five curves: The width, the position of the first pair of Gaussian peaks, the amplitude of the central peak, and the two amplitudes of the two pairs of Gaussian peaks respectively.

We find a standard deviation of $\sigma_{deriv} = 56.4$ (1.3) pA and thus derive a standard deviation of $\sigma_{direct} = 40$ (1) pA for the level broadening in the direct current data after background correction.

**SI-5 Theoretical models and results**

We model the RTS of the **{Co₄}**- and **{Mn₄}**-functionalised CNT quantum dot by a master equation visualised in figure S6a and b, respectively. The current levels (14 in Fig. 3(a) and three in Figs. 4 and 5(b) in the main paper) correspond to energies with equidistant spacing $\Delta\varepsilon$. We assume that each complex can be excited just once by this amount of energy. For level 0, all $K$ complexes are in their ground state. Each excitation of the molecules attached to the CNT leads to an increase of the level number by one. The main difference between the **{Co₄}** and the **{Mn₄}** systems is how the excitation by energy $\Delta\varepsilon$ is distributed among the grafted complexes. This strongly affects the different transition rates between the levels and, thus, the RTS.

For the **{Co₄}**-system, each excitation is localised at one complex only, i.e. the complexes are excited independently of each other. For $K = 70$ **{Co₄}** complexes there are $K = 70$ possibilities to choose the first complex to be excited (transition from level 0 to level 1) and $K - 1 = 69$ possibilities to choose the second (transition from level 1 to level 2).

For the relaxation from level 1 to level 0, there is no choice but one has to pick the given excited molecule, while for relaxation from level 2 to level 1 two excited molecules are available. In addition to these combinatory factors, there is a Boltzmann factor $r = e^{-\beta\Delta\varepsilon}$ for the excitation relative to the relaxation. The rate $\Gamma$ sets the overall time scale. This leads to the transition rates depicted in Fig. S6a.

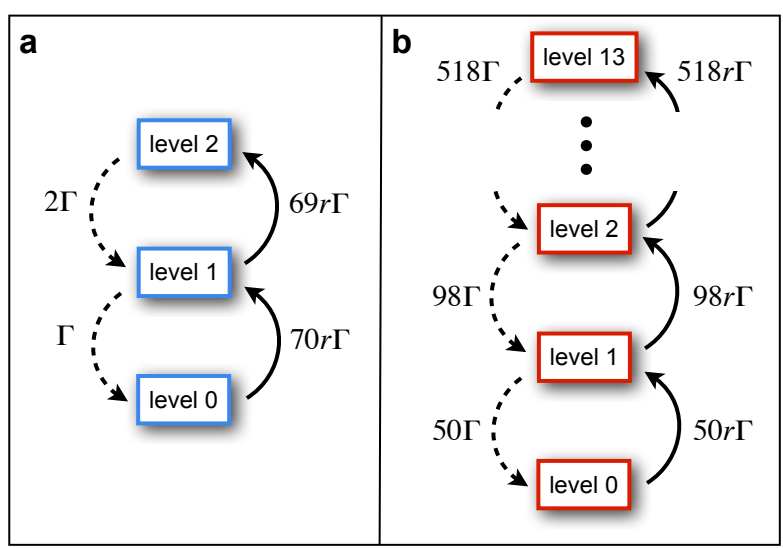

**Figure S6 | Sketch of the current levels and the transition rates. a,** the **{Co₄}**- and **b,** the **{Mn₄}**-functionalised CNT-quantum-dot device with transitions rate $\Gamma$ and a Boltzmann factor $r$. Dashed lines indicate the counted transitions.

For the {Mn$_4$}-system, we assume *collective* excitations involving *all* $K = 50$ {Mn$_4$} complexes. The excited states of the full system are coherent superpositions of the excitations of individual complexes. Beginning from the lowest level $|\psi_0\rangle = |0,0,0,\ldots,0\rangle$ with each complex in its ground state 0, we construct the states for the higher levels by applying the bosonic operator $\sum_{k=1}^{K} a_k^\dagger$, where $a_k^\dagger$ excites complex $k$ (but only if it is in its ground state), and by normalizing the state afterwards. We obtain

$$|\psi_n\rangle = \frac{1}{\sqrt{\binom{K}{n}}} \left(\sum_{k=1}^{K} a_k^\dagger\right)^n |0,0,0,\ldots,0\rangle, \qquad (4)$$

which leads to the matrix elements

$$\left|\langle\psi_{n+1}|\sum_{k=1}^{K} a_k^\dagger |\psi_n\rangle\right|^2 = (K-n)(n+1)$$

and, thus, to the rates depicted in Fig. S6b.

To calculate the factorial cumulants

$$C_{\mathrm{F},m}(\Delta t) = \langle\langle N(N-1)\ldots(N-m+1)\rangle\rangle = \partial_z^m \ln M_{\mathrm{F}}(z,\Delta t)|_{z=0}$$

for the {Co$_4$}-functionalised CNT quantum dot (and similar for {Mn$_4$}) assuming $K$ {Co$_4$} complexes, we write the master equation in the $N$-resolved form

$$\dot{P}_{N,0}(\Delta t) = -Kr\Gamma P_{N,0}(\Delta t) + P_{N-1,0}(\Delta t), \qquad (5)$$

$$\dot{P}_{N,1}(\Delta t) = Kr\Gamma P_{N,0}(\Delta t) - (1 + (K-1)r)\Gamma P_{N,1}(\Delta t) + 2\Gamma P_{N-1,2}(\Delta t), \qquad (6)$$

$$\dot{P}_{N,2}(\Delta t) = (K-1)r\Gamma P_{N,1}(\Delta t) - 2\Gamma P_{N,2}(\Delta t). \qquad (7)$$

Here, $P_{N,n}(\Delta t)$ is the probability that $N$ transitions have occurred and the system is in level $n$ at the end of the time interval $\Delta t$. After a $z$-transform,[14-15] we obtain the generating function $M_{\mathrm{F}}(z,\Delta t) = \vec{e}^{\,T} exp[\mathbf{W}_{z+1}\Delta t]\vec{P}_{\mathrm{stat}}$ with $\vec{e}^{\,T} = (1,1,1)$ and the stationary density matrix given by $\mathbf{W}_1\vec{P}_{\mathrm{stat}} = (0,0,0)^T$ and $\vec{e}^{\,T}\vec{P}_{\mathrm{stat}} = 1$. The matrix $\mathbf{W}_z$ is given by

$$\mathbf{W}_z = \begin{pmatrix} -Kr & z & 0 \\ Kr & -1-(K-1)r & 2z \\ 0 & (K-1)r & -2 \end{pmatrix}\Gamma. \qquad (8)$$

For the simulations presented in Fig. 7 of the main paper, the two parameters $\Gamma$ and $r$ have been determined such that the simulation reproduces the first factorial cumulant $C_{\mathrm{F},1}(10\mathrm{s})$ and the ratio $P_{\mathrm{occ}}(\psi_1)/P_{\mathrm{occ}}(\psi_0)$ of the occupations of level 1 and 0 exactly. Higher-order factorial cumulants and the probability of the higher levels are then a result of the simulation without any further fitting parameter.

In case of a system in which the counted transitions happen independently from each other, the sign of the $m$-th factorial cumulant must be $(-1)^{m-1}$ for all $\Delta t$.[16,15,17] This is, indeed, the case for independent two-level fluctuators, in accordance with the {Co$_4$}-data (blue curves in Fig. 7(b)-(d) of the main paper). The {Mn$_4$}-data (red curves in Fig. 7(b)-(d) of the main paper) show a different behaviour, i.e. they cannot be modelled by independent two-level fluctuators. The assumption of coherent superpositions of the excitations, on the other hand, can reproduce the data.

For completeness, we check yet another model for the **{Mn₄}**-data. Let us assume local excitations of independent *three-level* fluctuators, i.e. each complex can accommodate *two* excitations which, by chance, happen to have the same excitation energy. In Fig. S7, we depicted the results for this model (Fig. S7a). The additional parameter $\bar{\Gamma}$ is determined such that the simulation reproduces $C_{F,2}(10s)$ exactly. Indeed, the cumulants observed in the experiment can be reproduced by the simulation (Fig. S7c, d, and e). However, the occupation probabilities $P_{occ}$ are not recovered (Fig. S7b). Similar as for the two-level fluctuators, the combinatorial factors associated with the degeneracy of the levels strongly suppress the occupation probabilities in comparison to ones extracted from the measurements. Furthermore, very different values of $\bar{\Gamma}$ and $\Gamma$ would be needed without any physical justification. We, thus, conclude that only the model assuming coherent superpositions can reproduce the measured data.

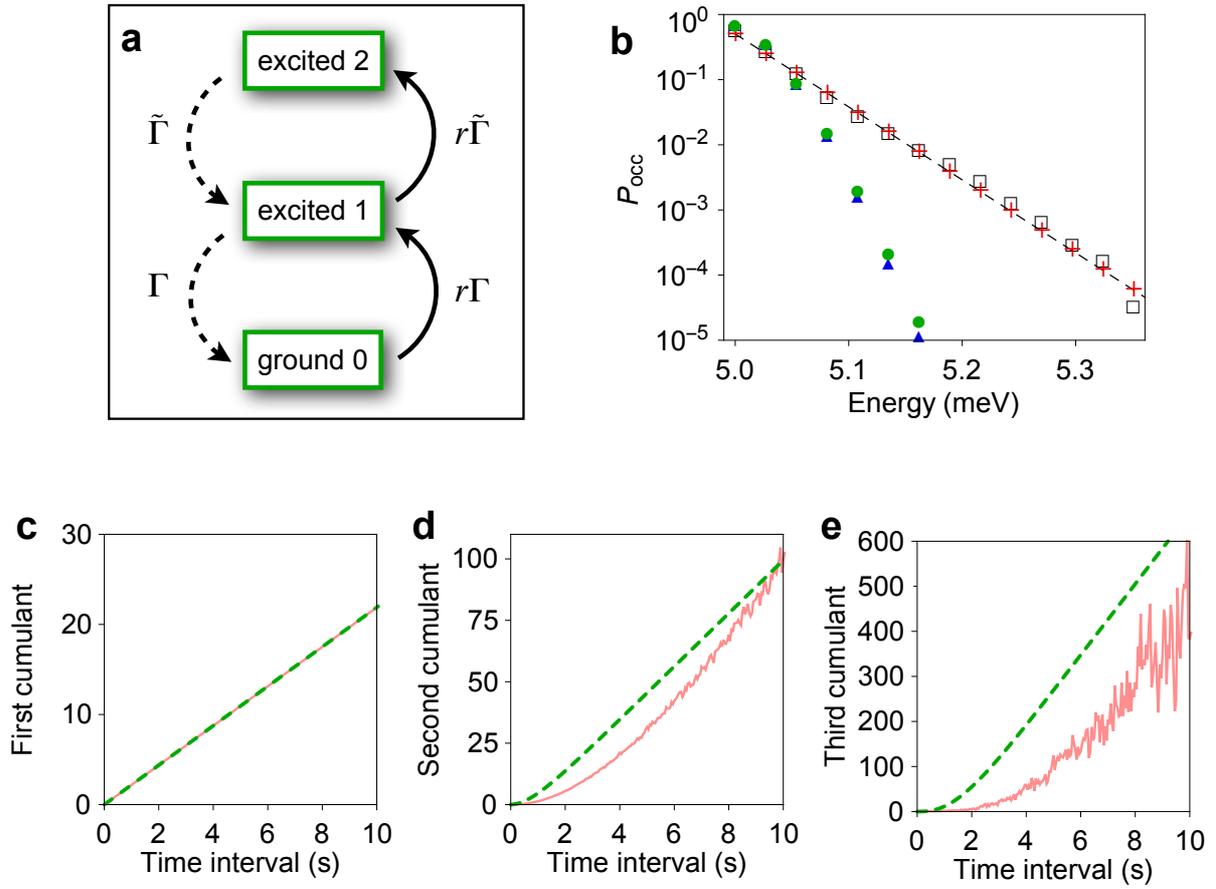

**Figure S7 | Sketch and modelling of a three-level fluctuator. a,** States and possible transitions of a three-level fluctuator. **b,** Occupation probability on a logarithmic scale for the different levels for the original data (black squares), and for models assuming $K = 50$ independent two-level fluctuators (blue triangles), $K = 50$ independent three-level fluctuators (green dots), and $K = 50$ coherently coupled two level fluctuators (red crosses). **c, d, e,** First, second, and third factorial cumulant as extracted from the measurements (red solid line) and a simulation assuming $K = 50$ independent three-level fluctuators (green dashed line) with $r = 0.01$, $\Gamma = 1.25 \ s^{-1}$, and $\tilde{\Gamma} = 316.99 \ s^{-1}$.